\documentclass[
               aps,
               prl,
               twocolumn,
               superscriptaddress,
               nofootinbib,
               showpacs]{revtex4-1}
\usepackage[utf8]{inputenc}
\usepackage{graphicx}

\usepackage{multirow}
\usepackage{array,booktabs}
\usepackage{hyperref}
\usepackage{amsmath}
\usepackage{amssymb}
\usepackage{amsfonts}
\usepackage{latexsym}
\usepackage{pifont}

\DeclareGraphicsExtensions{.eps}
\usepackage{color}

\usepackage{engord}
\newcommand{\twodots}[1]{\ensuremath{:\negmedspace #1\negmedspace:}}

\newcommand{\op}[2]{\hat{\textbf{#1}}_{#2}}
\newcommand{\dagop}[2]{\hat{\textbf{#1}}_{#2}^\dag}

\pacs{03.65.Wj, 42.50.Ar, 42.50.Dv, 42.50.-p}

\begin{document}

\title{Accessing higher order correlations by time-multiplexing}

\author{M.~Avenhaus}
\email[E-mail: ]{malte.avenhaus@mpl.mpg.de}
\affiliation{Max Planck Institute for the Science of Light, G\"unther-Scharowsky-Stra\ss{}e 1/Bau 24, 91058 Erlangen, Germany}
\author{K.~Laiho}
\affiliation{Max Planck Institute for the Science of Light, G\"unther-Scharowsky-Stra\ss{}e 1/Bau 24, 91058 Erlangen, Germany}
\author{M.~V.~Chekhova}
\affiliation{Max Planck Institute for the Science of Light, G\"unther-Scharowsky-Stra\ss{}e 1/Bau 24, 91058 Erlangen, Germany}
\affiliation{Department of Physics, M.~V.~Lomonosov Moscow State University, Leninskie Gory, 119992 Moscow, Russia}
\author{C.~Silberhorn}
\affiliation{Max Planck Institute for the Science of Light, G\"unther-Scharowsky-Stra\ss{}e 1/Bau 24, 91058 Erlangen, Germany}

\begin{abstract}
We experimentally measured higher order normalized correlation functions (nCF) of pulsed light with a time-multiplexing-detector. We demonstrate excellent performance of our device by verifying unity valued nCF up to the eighth order for coherent light, and factorial dependence of the nCF for pseudothermal light. We applied our measurement technique to a type-II parametric downconversion source to investigate mutual two-mode correlation properties and ascertain nonclassicality.
\end{abstract}

\keywords{quantum optics, photon state characterization}
\maketitle


Many optical effects involving coherence phenomena like interference, diffraction or radiation from fluctuating
sources can be related to the concept of mutual correlation \cite{Mandel}. Correlation functions unravel their full
potential in the theoretical framework of quantum optics, pioneered by the seminal work of Glauber
\cite{Glauber:Coherent}. The absence of second-order correlations in the famous antibunching experiment for light by
Kimble et. al. \cite{Kimble} has rigorously demonstrated the corpuscular theory of the photon for the first time.
Normally ordered correlation functions constitute a special case of arbitrary operators in quantum optics, e.g.
density matrices, which can be expressed by normally ordered boson operator moments \cite{Cahill:1969p8811:Wunsche:1990p8943}. Only recently a scheme has been proposed \cite{Shchukin:2006p918} to connect the measurement of
correlation functions with these generalized operator moments, rendering a new approach to quantum state
characterization.

The experimental study of correlation functions has a long standing history. Intensity correlation measurements
were first performed by Hanbury Brown and Twiss \cite{HanburyBrown:1956p8951} to determine the diameter of stars.
Coherent laser beams have been studied by Chopra et al. \cite{Chopra:1973p8340} who examined correlation
functions up to the third order. For parametric downconversion (PDC) correlations between signal and idler modes
have been analyzed up to the fourth order \cite{Ivanova:2006p8520} and nonclassical signs in second-order
correlation functions have been demonstrated for heralded single photons
\cite{Rarity:1987p1374:URen:2005p1553:Bussieres}, commonly employing multiple avalanche photodiodes (APD) as
detectors \cite{Boiko:2009p8960}. Another emerging topic is the phenomenon of ghost imaging, tightly related to
classical \cite{Scarcelli:2004p8944} versus quantum correlations \cite{Gatti:2004p9327}.

Though various experiments have been performed along the route of measuring normalized correlation functions (nCF) $g^{(n)}$ it remained challenging to measure to orders higher than four, especially for nonclassical state preparation. In this Letter we will present a scalable and integrated scheme to determine higher-order nCFs based on time multiplexed detection (TMD) \cite{Fitch2003:Achilles:2003p338} for pulsed sources. 

A normalized correlation function (nCF) $g^{(n)}$ in general is a time-dependent function of the electromagnetic
field or, in the context of quantum mechanics, of the creation and annihilation operators: $g^{(n)} = \frac{\langle
\dagop{a}{}(t_1) \cdots \dagop{a}{}(t_n) \op{a}{}(t_n) \cdots \op{a}(t_1) \rangle} {\langle \dagop{a}{}(t_1)
\op{a}{}(t_1) \rangle \cdots \langle \dagop{a}{}(t_n) \op{a}{}(t_n) \rangle}$. This definition of normalization renders $g^{(n)}$ for monochromatic states time-independent, a finding which remarkably also applies for
single-mode pulsed ensemble measurements. In fact, any time-dependence of $g^{(n)}$ can be directly related to
multimodeness, such that its measurement yields information for distinguishing single-mode from multimode
states. Conventionally, the time dependence of nCF is probed in the regime of CW light with help of detectors
much faster than the amplitude fluctuations. In opposition, we focus here on the measurement of $g^{(n)}$ for
pulsed sources \cite{Ivanova:2006p8520} with a pulse envelope much shorter than the detector response time. An
intuitive physical interpretation is that the integration over a deltalike intensity pulse essentially samples
the nCF only at the maximum of the pulse envelope. Any multimodeness can then either be described as a temporal
substructure within the nCF of the pulse, or more conveniently in an effective single-mode picture: The photon
number statistics, for instance, of a thermal light source converges to a Poissonian distribution with increasing
number of modes, and is caused by a convolution effect \cite{Avenhaus:2008p6044}. This comes along with a
decrease of $g^{(n)}$ for the described scenario. Note, however, that this decrease is not due to temporal
smoothing but in our context of ultrashort pulses is only caused by the multimode character.

Other features of $g^{(n)}$ also provide an easy way for classifying quantum states: The second-order nCF 
is a unique operational measure for non-classicality \cite{Klyshko:1996p4925} if $g^{(2)}(0) < 1$. For various
applications like quantum key distribution or Schrödinger cat state preparation, great interest lies in the
feasibility to prepare $m$-photon states \cite{OSullivan:2008p6134}. Here, correlation functions can be used to
quantify the fidelity of an $m$-photon state since all $g^{(n)} = \frac{m (m-1) \dots (m-n+1)}{m^n}$ must be zero for
$n>m$. Lately, a profound and highly applicable theoretical framework has been introduced \cite{Vogel:2008p3453} to
distinguish classical radiation from nonclassical one via correlation properties. For experimentalists an intriguing
feature of a nCF is its loss \emph{independence}, i.e no a priori assumptions about channel or detector efficiency are required.

\begin{figure}[htbp]
    \centering
        \includegraphics[width=0.9\linewidth]{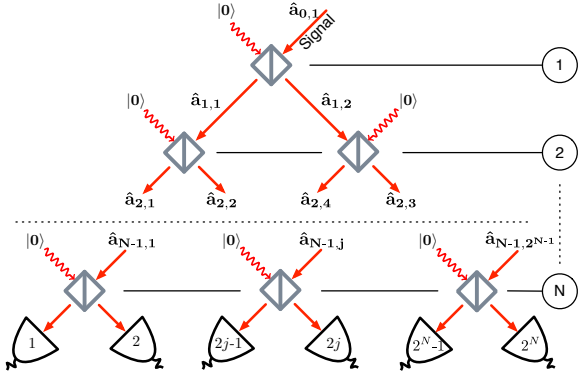}
    \caption{Spatial mode multiport: Beamsplitter cascade mixing signal input with vacuum at each beamsplitter}
    \label{fig:BSC-Spatial}
\end{figure}

\label{sec:theory}
The theoretical model of our experimental setup corresponds to a hierarchical beamsplitter cascade of $N$ stages as shown in Fig.~\ref{fig:BSC-Spatial}. Losses can be accounted for by adding a virtual beamsplitter in front of each detector, such that only the transmitted amplitude is sensed. In the following we mathematically elucidate this experimental configuration to be harnessed for measuring $g^{(n)}$ up to $n=2^N$.
                                                                                                          
Suppose that each beam splitter at stage $k$ transforms the input mode $\op{a}{k-1,j}$ and vacuum mode $\op{b}{k-1,j}$ (not shown) into $\op{a}{k,2j-1}$ and $\op{a}{k,2j}$ as illustrated in Fig.~\ref{fig:BSC-Spatial}. We introduce the parameter $S_{k,l}$ in the transformation for a beamsplitter, denoting a transmission amplitude for the second index $l$ being odd, and reflection amplitude for $l$ being even:
\begin{equation}
    \begin{array}{@{}lclcl@{}}
	    \op{a}{k,2j-1} & = & S_{k,2j-1}\op{a}{k-1,j} & - & S_{k,2j}\op{b}{k-1,j} \\
	    \op{a}{k,2j} & = &  S_{k,2j}\op{a}{k-1,j} & + & S_{k,2j-1}\op{b}{k-1,j}\\   
    \end{array}
	\label{eq:bogoliubov}
\end{equation}
Starting from mode $\op{a}{0,1}$ the signal transverses the cascade to a marked detector $j$ along a unique
path $\mathcal{P}_j^0 \rightarrow \mathcal{P}_j^1 \rightarrow \cdots \rightarrow \mathcal{P}_j^N$. Each
mode along this route is denoted as $\op{a}{P_j^k}$. The intensity correlation $\langle \dagop{a}{N,j_1}
\op{a}{N,j_1} \cdots \dagop{a}{N, j_n} \op{a}{N, j_n} \rangle$ between $n$ distinct and independent detector modes can be reordered to read $\langle \dagop{a}{N,j_1} \cdots \dagop{a}{N, j_n} \op{a}{N, j_n} \cdots \op{a}{N,j_1} \rangle$.
This expectation value can be stepwise expressed by modes of the previous beamsplitter stage with help of
Eq.~\ref{eq:bogoliubov}. By explicitly tracing out the vacuum modes and, iterating back to the origin, we
obtain:
\begin{widetext}
\[    \langle \dagop{a}{\mathcal{P}_{j_1}^N} \cdots \dagop{a}{\mathcal{P}_{j_n}^N} \op{a}{\mathcal{P}_{j_n}^N} \cdots \op{a}{\mathcal{P}_{j_1}^N} \rangle =
  \langle 
  S^\ast_{\mathcal{P}_{j_1}^N} \dagop{a}{\mathcal{P}_{j_1}^{N-1}} \cdots 
  S^\ast_{\mathcal{P}_{j_n}^N} \dagop{a}{\mathcal{P}_{j_n}^{N-1}}  
  S_{\mathcal{P}_{j_n}^N} \op{a}{\mathcal{P}_{j_n}^{N-1}} \cdots
  S_{\mathcal{P}_{j_1}^N} \op{a}{\mathcal{P}_{j_1}^{N-1}} 
  \rangle = 
  \left( \prod_{k=1}^{N} \prod_{l=1}^{m} |S_{\mathcal{P}_{j_l}^k}|^2 \right)
  \langle \hat{\textbf{a}}^{\dagger m}_{0,1} \hat{\textbf{a}}^m_{0,1} \rangle
\]
\end{widetext}
The value of $g^{(n)}$ for the signal of interest can therefore be obtained when dividing the mutual intensity
correlation of $n$ detectors divided by the intensity of each individual detector. All losses, transmission and reflection amplitudes resembled by the product of $S_{l,k}$ factorize in both numerator and denominator and thus cancel out. Hence, our calculations confirm $g^{(n)}$ to be loss independent.
\begin{equation}
     g^{(m)} = \frac{\langle
    \hat{\textbf{a}}^{\dagger m}_{0,1} \hat{\textbf{a}}^m_{0,1} \rangle} {\langle
    \hat{\textbf{a}}^{\dagger}_{0,1} \hat{\textbf{a}}_{0,1} \rangle^m} =
    \frac{\langle \dagop{a}{N,j_1} \op{a}{N,j_1} \cdots \dagop{a}{N, j_n}
    \op{a}{N, j_n} \rangle} {\langle \dagop{a}{N,j_1} \op{a}{N,j_1} \rangle \cdots
    \langle \dagop{a}{N, j_n} \op{a}{N, j_n} \rangle}
    \label{eq:gn}
\end{equation}
Note that this result can easily be generalized to 
cross-correlate $g^{(m,n)}$ for two spatial modes $\op{a}{}$ and $\op{b}{}$ with independent cascades by considering coincidences of $m$ detectors in mode $\op{a}{}$ with $n$ detectors in mode $\op{b}{}$, followed by an analogous normalization.

The measurement of $g^{(n)}$ yet requires linear detectors with respect to the intensity $\op{n}{} =
\dagop{a}{}\op{a}{}$. Using linear photodiodes would pose a straightforward approach. However, their signal
at the single photon level is concealed by the readout and amplification noise. Conventionally, APDs are
employed that operate in Geiger mode but only yield a binary click response upon the detection of photons.
Since a detection process causes an avalanche in the detector's band gap, the number of impinging
photons cannot be resolved. On the other hand, the \emph{probability} for a click is linear under certain
conditions. The probability to excite $m$ photoelectrons is given by the operator $ \op{P}{m} =
\;\twodots{\frac{1}{m!}(\eta \op{n}{})^m \exp(-\eta \op{n}{})} $, with detection efficiency $\eta$ and
$\twodots{\;\;}$ denoting normal ordering. Since our detector cannot discriminate the number of
photoelectrons, any number $m$ greater than one $\op{P}{} \equiv \sum_{m=1}^\infty \op{P}{m}$ gives rise to a
click. For the regime of states with low click probability, $\op{P}{} \approx \eta \op{n}{}$ is dominated by
the linear term. In the limit of low click probability an APD acts as a linear but lossy detector
which can be employed to measure the loss invariant nCF in Eq. \ref{eq:gn}. Neglecting dark counts, the
approximation to the linear behavior of $\op{P}{}$ becomes even better for higher losses. The accuracy of
the correlation measurement then increases, however at the cost of statistical precision.
                                        
\begin{figure}[tbp]
    \centering
        \includegraphics[width=1.0\linewidth]{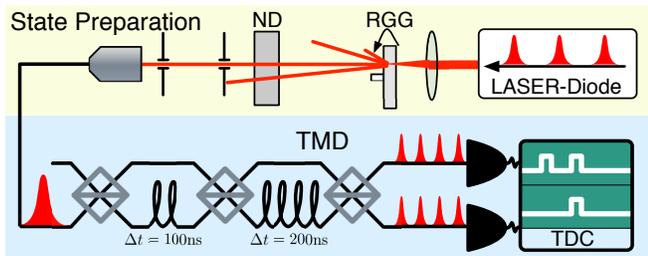}
    \caption{Time mode multiport: Experiment for correlation measurements of coherent and chaotic states by a TMD. For abbreviations see text.}
    \label{fig:BSC-Temporal}
\end{figure}

\label{sec:tmd}
The implementation of the free space spatial mode multiport in Fig.~\ref{fig:BSC-Spatial} even for a modest amount of
stages becomes very resource consuming. The same network, however, has previously been realized as a stable fiber
integrated TMD \cite{Fitch2003:Achilles:2003p338}, depicted in Fig.~\ref{fig:BSC-Temporal}. Here, the spatial
modes of the cascade are mapped to modes in the time domain. The previous theoretical considerations therefore remain
identical. One additional fiber loop between two subsequent beamsplitters causes a delay between the temporal modes,
which has been chosen to be twice the APD dead time.
\label{sec:test_scenario}

Coherent and thermal states provide an ideal testbed for nCFs. Firstly, they are easy to prepare, and secondly, they have a simple analytic expression: $g^{(n)} = 1$ for coherent and $g^{(n)} = n!$ for thermal states. Coherent states are readily provided by a laser source whereas pseudo thermal sources can be mimicked by chaotic states, easily obtained from a moving speckle pattern \cite{Arecchi:1965p8701}. We employed the coherent state and the chaotic state in order to validate the TMD's potential for measuring nCFs.  

Our first experiment was conducted to demonstrate the capability of measuring $g^{(n)}$ with a TMD, and the
experimental configuration is depicted in Fig.~\ref{fig:BSC-Temporal}: A diode laser generated 50\,ps pulses of
single-mode coherent states at 810\,nm with a rate of 20\,kHz. The pulse was heavily attenuated by a neutral density
(ND) filter to the single photon level. The photons were coupled into one input of a multimode fiber based TMD. This
yielded an average count rate of 3.3\,kHz per detector mode. The electronic detector signal was processed by a
time-to-digital converter (TDC). The TDC numbered consecutively all laser pulses and recorded the time information of
an APD-click with a precision of 81\,ps relative to the laser trigger, much better than the integration time of the
APD (1\,ns). The chaotic light was generated by focussing the laser onto a rotating ground glass (RGG). This way a
moving speckle pattern was formed and only an areal portion of less than an average speckle was coupled into a
single-mode fiber based TMD with help of two additional apertures (AP). Here, the laser repetition rate was increased
to 500\,kHz and an average count rate of 6.6\,kHz observed per TMD mode.

Extracting $g^{(n)}$ from the recorded data was performed by selecting $n$ arbitrary TMD modes. Let all those
possible ${2^N \choose n}$ selections of $n$ modes be contained in set $\mathcal{C}$. For the particular case of
$n=2$ the set $\mathcal{C}$ consists of all pairs, exemplarily shown in Table~\ref{tab:g2_subsets} for a $g^{(2)}$
measurement of a coherent input state. Each combination $c \in \mathcal{C}$ provides an approximation to $g^{(n)}$ when dividing the $n$-fold coincidence probability by the product of individual single click probabilities according
to Eq.~\ref{eq:gn}. This gives a string of values where the mean reaches a value closest to the real $g^{(n)}$. Some
benefits are offered when a coincidence measurement does not select all $2^N$ modes and $\mathcal{C}$ contains
multiple elements. Firstly, part of the signal is diverted to other modes not being part of the particular selection.
Hence, measurement accuracy is increased due to higher losses considering $c$ alone. Secondly, the signal is not lost
within the \emph{entire} multiport network but is transferred to other combinations in $\mathcal{C}$. Therefore, good
measurement statistics can be maintained by averaging over the results of $\mathcal{C}$. Additionally, a rough error
estimation can be given by the standard deviation.

Our results for higher order correlations are presented in Table~\ref{tab:gn}. We find an excellent agreement between
theoretical prediction and experimental findings for both coherent and chaotic states. We have proven with confidence
that the TMD can be utilized for accurate measurement of nCFs and provides the basis for more advanced quantum state
characterization. Interestingly we find an overestimation of the error for $n \gtrsim 5$. Contribution to the growing standard deviation originates from an increasing number of possible combination in $\mathcal C$. Another, more fundamental effect is the growth in the standard deviation of $g^{(n)}$ for chaotic light
and large $n$ as a result of strong intensity fluctuations. It is this growth that reduces
the signal-to-noise ratio of multiphoton ghost imaging with chaotic light sources \cite{Chen:2009p9292}.

\begin{figure}[htbp]
    \centering
    \includegraphics[width=0.99\linewidth]{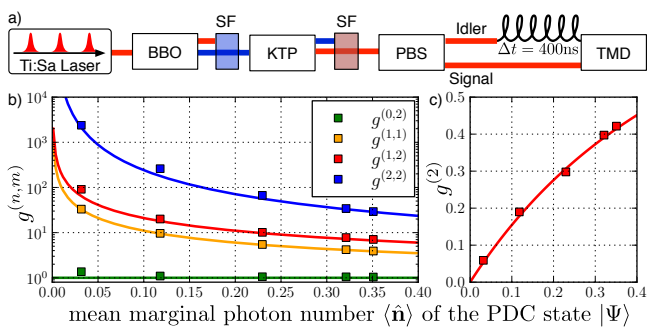}
    \caption{a) Experimental setup for mutual correlation measurements. b) shows the dependency of mutual correlations $g^{(n,m)}$ in theory (solid) and measurement (squares) for five different $\langle \op{n}{} \rangle$. c) displays $g^{(2)}$ of the idler beam constrained on an APD click in the signal beam, hence demonstrating the fidelity of heralding a single photon state.}
    \label{fig:InterTMD}
\end{figure}

\begin{table}
\begin{tabular}{ll|ccccccc}
 &       & \multicolumn{7}{c}{TMD mode} \\
 & $g^{(2)}$ & 2 & 3 & 4 & 5 & 6 & 7 & 8\\ \hline
 \multirow{7}{*}{\rotatebox{90}{TMD mode}}
 & 1 & 1.00 & 1.00 & 0.99 & 1.01 & 1.00 & 1.01 & 1.00 \\
 & 2 & -- & 0.99 & 1.00 & 1.00 & 1.00 & 1.00 & 1.01 \\
 & 3 & -- & -- & 1.00 & 1.00 & 0.99 & 1.01 & 1.02 \\
 & 4 & -- & -- & -- & 1.00 & 1.01 & 1.00 & 1.00 \\
 & 5 & -- & -- & -- & -- & 1.00 & 1.00 & 1.00 \\
 & 6 & -- & -- & -- & -- & -- & 1.00 & 0.99 \\
 & 7 & -- & -- & -- & -- & -- & -- & 1.01 \\ \hline 
\end{tabular}
\caption{All 28 possible second-order coincidence subsets for eight total TMD modes. Displayed are the measurement results for $g^{(2)}$ after the first 30\,s of measurement time for the coherent light scenario, yielding a mean $g^{(2)} = 1.002 \pm 0.007$.}
\label{tab:g2_subsets}
\end{table}
{
\begin{table}
\renewcommand{\arraystretch}{1.2}
\begin{tabular}{crrrrrrrll}
	& $g^{(2)}$ & $g^{(3)}$ & $g^{(4)}$ & $g^{(5)}$ & $g^{(6)}$ & $g^{(7)}$ & $g^{(8)}$ & & \\ \hline
    \multirow{3}{*}{\rotatebox{90}{Coherent}}
    & 1.0000 & 1.0000 & 1.0000 & 1.0000 & 1.0000 & 1.0000 & 1.0000 & \vline & Th.\\
    & 1.0006 & 1.0012 & 1.0016 & 1.0021 & 1.0032 & 1.0045 & 0.9871 & \vline & Exp.\\
    & 0.0010 & 0.0022 & 0.0042 & 0.0105 & 0.0298 & 0.0547 & --- & \vline & Err.\\ \hline \hline
	\multirow{3}{*}{\rotatebox{90}{Chaotic}}
	& 2.00 & 6.00 & 24.00 & 120.00 & 720.00 & 5040 & 40320 & \vline & Th. \\
	& 2.03 & 6.28 & 25.23 & 120.23 & 651.54 & --- & --- & \vline & Exp. \\
	& 0.01 & 0.12 &  2.30 &  44.69 & 941.10 & --- & --- & \vline & Err. \\ \hline
\end{tabular}
\caption{Measurements of $g^{(n)}$ for coherent and chaotic light sources. Displayed are the theoretical prediction (Th.) in the \engordnumber{1}, the experimental value (Exp.) in the \engordnumber{2}, and the error estimation (Err.) in the \engordnumber{3} row.}
\label{tab:gn}
\end{table}
}
\label{sec:application}
In our second experiment, depicted in Fig. \ref{fig:InterTMD}a, we investigated correlations and
non-classicality between signal and idler beams from the nonlinear process of parametric downconversion
(PDC). Since the photon number between signal and idler strictly match for a PDC state
$|\Psi\rangle = \sum_n \sqrt{p_n} |n,n\rangle$, the measurement of nCF is particularly interesting. The beam
from an ultrafast (100\,fs) Ti:Sa laser system, pulsed at 1\,MHz at 796\,nm) was frequency doubled in a
nonlinear $\beta\text{-BaB}_2\text{O}_4$ (BBO) crystal and used to pump a PDC process (order of 100\,fs) in a periodically poled $\text{KTiOPO}_4$ (KTP) waveguide chip.
Any residual pump light was eliminated by spectral filters (SF). Signal and idler modes emerged from the
type-II PDC process with orthogonal polarization and were spatially separated by a polarizing beamsplitter.
Both beams were launched into input fibers of our TMD, one being delayed by 400\,ns to emulate two
independent TMDs.


We parametrized the mutual correlation function $g^{(n,m)} = \frac{\langle \hat{\textbf{a}}^{\dagger n}
\hat{\textbf{a}}^{n} \hat{\textbf{b}}^{\dagger m} \hat{\textbf{b}}^{m} \rangle}{\langle \hat{\textbf{a}}^{\dagger}
\hat{\textbf{a}} \rangle^n \langle \hat{\textbf{b}}^{\dagger} \hat{\textbf{b}} \rangle^m}$ in terms of the mean
photon number for comparing measurement and theory. The model assumed our PDC state to exhibit a Poissonian
distribution $p_n$ due to the excitation of many spectral modes \cite{Avenhaus:2008p6044}. The experiment was
conducted with five different sets of pump power resulting in different mean marginal photon numbers $\langle
\op{n}{} \rangle$. In order to calibrate $\langle \op{n}{} \rangle$ we applied a loss estimation procedure for TMDs
as presented in \cite{Worsley:2009p8528}. The results for $g^{(0,2)}$, $g^{(1,1)}$, $g^{(1,2)}$ and $g^{(2,2)}$ are
presented in Fig.~\ref{fig:InterTMD}b. According to \cite{Vogel:2008p3453} a possible criteria for the violation of
classicality is $g^{(1,2)} = \gamma \sqrt{g^{(2,2)} g^{(0,2)}}$ for $\gamma>1$. Our experimental data clearly affirms this violation with a value of $\gamma$ ranging between 1.19 and 1.60, depending on pump power. We therefore can testify non-classicality of $|\Psi\rangle$, even for lossy detection. Heralding a single photon state upon a click in
one mode can leave remnants of higher order photon contributions in the other mode. This can be seen in
Fig.~\ref{fig:InterTMD}c for a conditioned $g^{(2)}$ measurement, proving nCFs to be an extremely powerful tool for
analyzing state preparation techniques. Note the excellent agreement between measurement and theory without
\emph{any} fit parameter, confirming the potential of the TMD as well as the Poisson statistics obtained from
multimode PDC.




\label{sec:so_long_and_thanks_for_all_the_fish} To conclude, we have analyzed in theory and
experiment the use of a beamsplitter cascade for measuring normalized correlations functions
$g^{(n)}$ with standard APD detection. We have implemented the cascade as an integrated and
easily scalable TMD fiber network. The feasibility to correctly measure high order $g^{(n)}$ has
first been ascertained by examining coherent and pseudothermal states of light. After this, we
applied our measurement method to PDC states to study two-mode correlation functions, the
non-classicality of PDC states, as well as $g^{(2)}$ for heralded single photon states with
remnant higher photon number contributions. We consider our experiments to be a first step paving
a new way for state characterization with TMDs utilizing arbitrary moments \cite{Shchukin:2006p918}, which is capable
of detecting quantum features of correlations independent of detrimental loss effects.


We would like to thank I.~N.~Agafonov and W.~Vogel for fruitful discussions. This work was supported by the EU under QAP funded by the IST directorate as Contract no. 015848.

\label{sec:references}



\end{document}